# Implicit neural representation for free-breathing MR fingerprinting (INR-MRF): co-registered 3D whole-liver water T1, water T2, proton density fat fraction, and R2* mapping

Chao Li[1,2], Jiahao Li[1,3], Jinwei Zhang[1,3], Eddy Solomon[1], Alexey V. Dimov[1], Pascal Spincemaille[1], Thanh D. Nguyen[1], Martin R. Prince[1] and Yi Wang[1]

1. Radiology, Weill Cornell Medicine, New York, NY, United States
2. Applied and Engineering Physics, Cornell University, Ithaca, NY, United States
3. Meinig School of Biomedical Engineering, Cornell University, Ithaca, NY, United States

## Abstract:

**Purpose**: To develop an MRI technique for free-breathing 3D whole-liver quantification of water T1, water T2, proton density fat fraction (PDFF), R2*.

**Methods**: An Eight-echo spoiled gradient echo pulse sequence with spiral readout was developed by interleaving inversion recovery and T2 magnetization preparation. We propose a neural network based on a 4D and a 3D implicit neural representation (INR) which simultaneously learns the motion deformation fields and the static reference frame MRI subspace images respectively. Water and fat singular images were separated during network training, with no need of performing retrospective water-fat separation. T1, T2, R2* and proton density fat fraction (PDFF) produced by the proposed method were validated in vivo on 10 healthy subjects, using quantitative maps generated from conventional scans as reference.

**Results**: Our results showed minimal bias and narrow 95% limits of agreement on T1, T2, R2* and PDFF values in the liver compared to conventional breath-holding scans.

**Conclusions**: INR-MRF enabled co-registered 3D whole liver T1, T2, R2* and PDFF mapping in a single free-breathing scan.

# 1 | Introduction

Chronic liver disease affects 1.5 billion people globally, resulting in two million deaths each year.(1, 2) The leading cause is nonalcoholic fatty liver disease (NAFLD).(3) The progressive form of NAFLD, known as nonalcoholic steatohepatitis (NASH), is often accompanied by liver inflammation and fibrosis.(3) Liver fibrosis can progress to cirrhosis, an end-stage condition that is fatal.(4, 5) In chronic liver disease, fibrosis, iron, and fat are interrelated and iron may be both a cause(6) and consequence of liver disease.(7-9)

Proton density fat fraction (PDFF) is a widely used MRI marker for measuring fat content in tissues,(10-12) while R2* is used to assess liver iron levels.(13, 14) Although liver biopsy remains the gold standard for diagnosing liver fibrosis, it's invasive, not suitable for regular monitoring, prone to variability between observers ,and has the risk of complications. Lately, more attention has been given to the potential of T1 and T2 measurements in assessing liver fibrosis and inflammation. Previous studies have shown that T1 relaxation time is significantly increased, and T2 relaxation time is notably reduced in fibrotic and cirrhotic livers,(13, 15-18)suggesting that these measurements could serve as non-invasive biomarkers for diagnosing liver disease and monitoring it over time.

However, there are still technical challenges that limit the use of MRI for liver diseases. While PDFF and R2* can be measured together in a single 3D scan, T1 and T2 usually require multiple 2D scans. The most common sequences for these measurements, 2D MOLLI for T1(19) and 2D multi-echo spin echo (MESE) for T2(20), require multiple breath-holds to capture different slices. This leads to longer scan times, discomfort for patients, and misregistration between slices.

In recent years, magnetic resonance fingerprinting (MRF)(21) has gained attention as a method for simultaneous multi-parametric mapping.(22, 23) By continuously sampling the magnetization during scanning, MRF can encode several parameters at once, generating co-registered, multi-parametric maps from a single scan. However, the k-space data for these signals is highly under-sampled, which can cause errors in the parametric maps.(24) This issue becomes more complex in free-breathing scans, where motion can introduce artifacts if not carefully addressed. Significant efforts have been made to develop algorithms that improve the reconstruction of under-sampled k-space data, including model-based iterative methods and learning-based approaches. For example, HD-PROST reconstructs multi-contrast MR images by using redundant information at both local and non-local levels and exploiting the correlations between different contrasts.(25) Recently, a deep learning framework was developed to generate high-resolution, co-registered T1, T2, R2*, and QSM maps for the brain by leveraging a network-designed k-space sampling pattern.(26) For free-breathing liver scans, MR-multitasking was introduced to create 3D T1, R2*, and PDFF maps,(27) and HD-PROST with motion compensation was used to generate 3D T1 and T2 maps.(22)

Recently, a new machine learning technique called implicit neural representation (INR) has shown potential in medical image reconstruction, registration, and analysis.(28-32) In INR, a neural network maps spatial coordinates to desired physical values, such as MRI signal intensities at specific voxels, by implicitly learning the geometry of the physical subjects.(33, 34) This method has recently been applied to dynamic image reconstruction using a deformation-driven technique.(31) However, to fully eliminate motion artifacts, this approach typically requires principal components of motion deformation fields, which are usually calculated offline from an external dataset. Using motion fields obtained with the data itself can lead to artifacts, as non-deformation intensity changes between the reference image and images from other motion states can be caused by aliasing or motion artifacts.(31)

In this paper, we demonstrate the feasibility of using INR to simultaneously learn motion deformation fields and reference frame images from scratch and in zero-shot, without any prior knowledge of the motion deformation fields. We show that this approach is effective in MRF, where navigator signals and images are acquired in transient states. In summary, we propose a pulse sequence and INR-based reconstruction method to create 3D whole-liver co-registered T1, T2, R2*, and PDFF maps simultaneously in a free-breathing scan.

# 2 | Methods

## 2.1 | Pulse sequence

The multi-contrast pulse sequence proposed, illustrated in Figure 1(a), draws inspiration from mcLARO(26) and MR multi-tasking (27)designs. It consists of a module that begins with a non-selective inversion pulse which is used to sensitize T1 relaxation, followed by 6 segments each consisting of a navigator pulse followed by 15 spiral multi-echo gradient-echo spiral readouts for T2* measurement. The navigator pulse, a central k-space line in the SI direction (kx = ky = 0), is played to capture the respiratory motion. In each segment, the readout pulses have the same angle with kz reordering sampled from a Gaussian distribution. The spiral leaves rotate at a golden angle between each segment, as shown in figure 1(b). Another module including a T2prep pulse (41,42) followed by the same 6 navigator-readout segments was used for T2 relaxation measurement, leading to a total of 12 readout segments in each repetition.

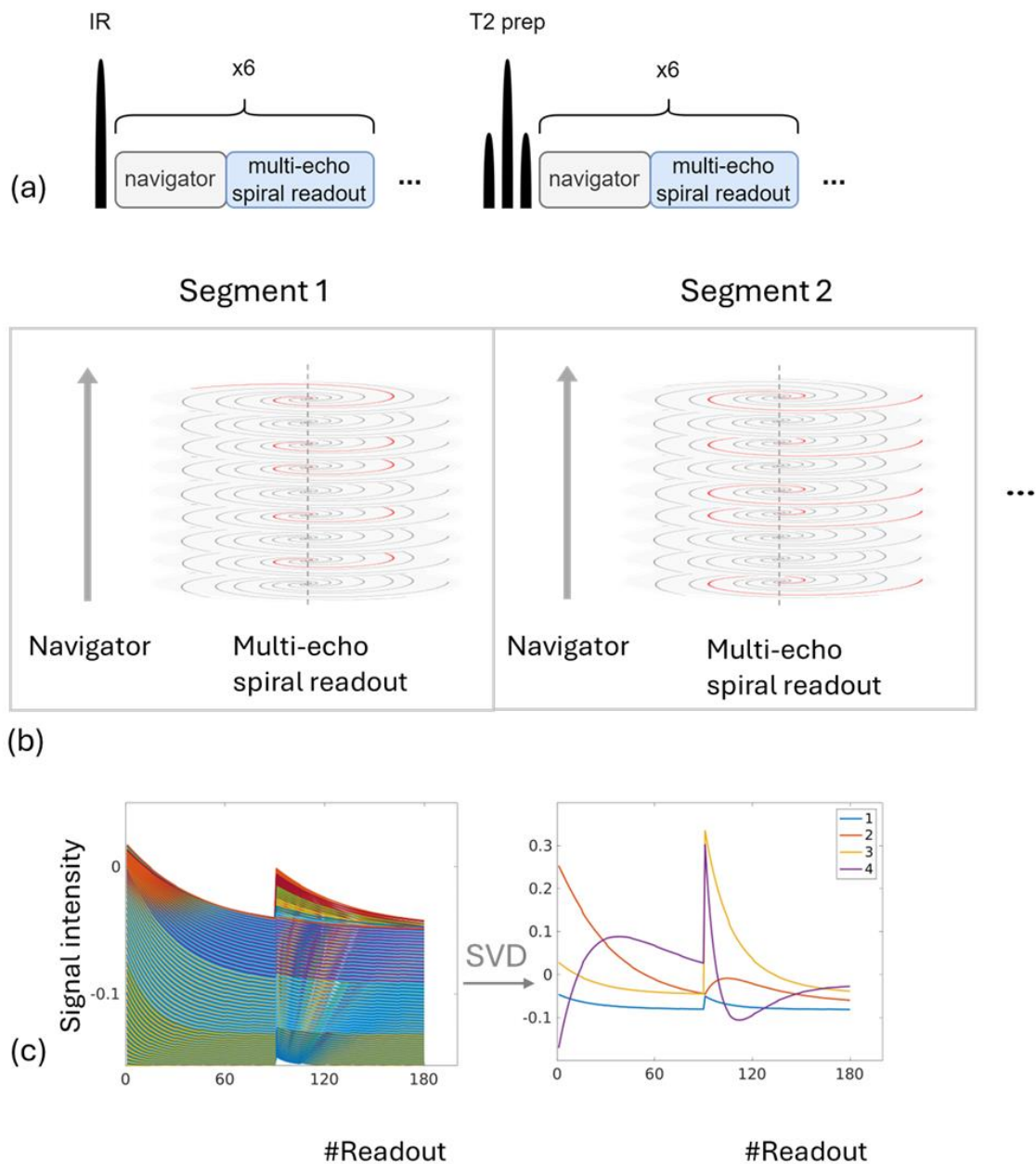

Figure 1: (a) Schematics of the proposed INR-MRF pulse sequence which consists of inversion recovery (IR) and T2prep magnetization preparations and multi-echo GRE readout segments; (b) A detailed view of the navigator pulse and the multi-echo spiral readout segments. A Cartesian navigator pulse with readout along kz was followed by a partial stack of spiral readouts in kx-ky plane with randomly sampled-kz. Acquired spiral leaves were indicated by red. The spiral leaves rotate a golden angle between each segment. (c) Bloch simulation of the steady state signals in the dictionary. A singular value decomposition was performed to compress the dictionary into 4 principal components $U_r$.

## 2.2 | Motion signal as input of the neural network

Because the magnetization is in transient state within each repetition, motion-binning using principal component-based method(35) designed for steady state imaging led to problem. Here we propose a simple envelope detection method that bins the respiratory motion for navigator signals in transient states. A main coil located near the diaphragm was manually selected, the navigator signal intensity from this coil was normalized, as shown in figure 2.a, and the local maxima are then detected to locate the position of the liver-lung boundary which provides the information of respiratory state (red curve in figure 2.a).

The motion signal ($m_n$ in figure 3) as the input of the 4D INR was constructed as follows: The position of local maxima of the navigator signal described previously were normalized to 0 and 1(figure 2.b), where 0 and 1 denote the end-of expiration and end-of-inspiration respectively. The points in this normalized navigator signal curve were classified into 12 groups based on which segment they belong to in their corresponding repetition (figure 2.c). For example, as shown in figure 2.c, circles denote the navigator signals in the first acquisition segment in the pulse sequence, and right triangles indicate the navigator signals in the last segment.

Navigator signals were sorted in ascending order within each of the 12 groups, and a sliding window was applied to each group. This window always encloses r navigator signals. It slides from end-of expiration and end-of-inspiration as from 0 to 1. Every time the window moves up, it skips j navigator signals (j<r), leading to r-j shared views between window n and n+1 (figure 2.c). If the whole scan contains R repetition of the pulse sequence, then there are a total of $\left\lfloor \frac{R-r+j}{j} \right\rfloor$ possible window positions, n = 1,2,... $\left\lfloor \frac{R-r+j}{j} \right\rfloor$. In this work, we chose the jump size j to be 1 and window size r to be 24. Assume the average navigator signal in window n for segment i was calculated to be $\overline{m_{n,i}}$. Motion representation $m_n$ was obtained by further average $\overline{m_{n,i}}$ over all segments (i=1..12):

$$m_n = \frac{1}{N} \sum_{i=1}^{N} \overline{m_{n,i}} \quad (1)$$

where N = 12 is the number of acquisition segments for our pulse sequence design. This number, as a representation of the respiratory state, will be used as the input of the neural network.

When training the network, a random window index n was chosen from 1,2,… $\left\lfloor \frac{R-r+j}{j} \right\rfloor$ for each iteration, and only the k-space readouts associated with the navigators bound by sliding window n in the 12 segments were fed into the network in this iteration. The readouts in one segment have similar contrasts, and thus our segment-selective (or contrast-selective) motion binning strategy ensures that there are same number of readouts in each contrasts being seen by the network in one iteration.

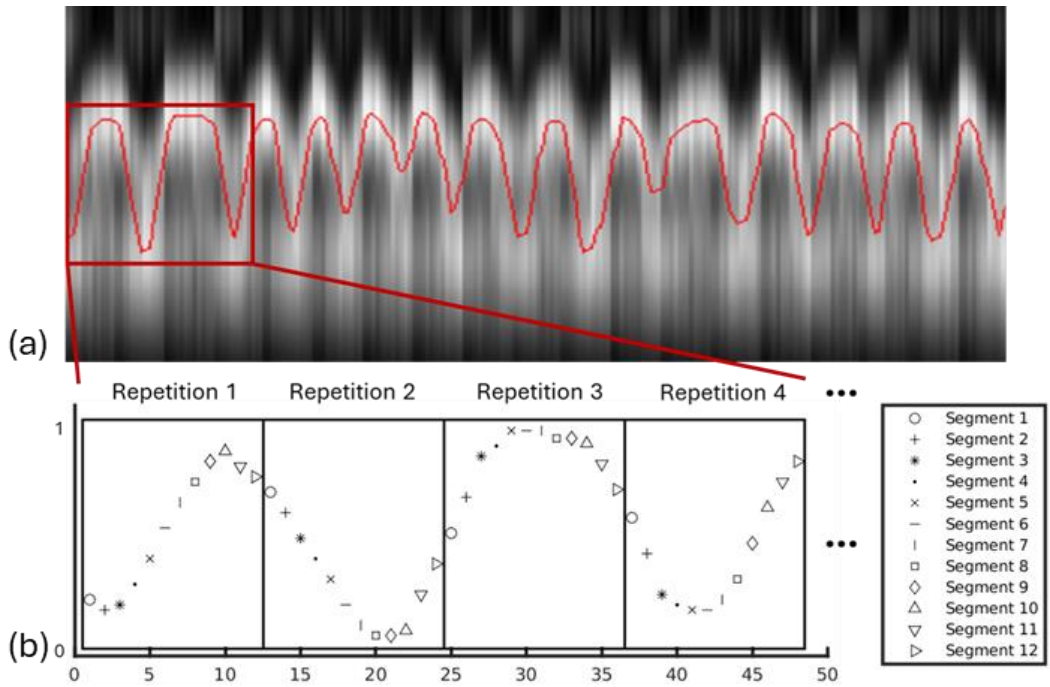

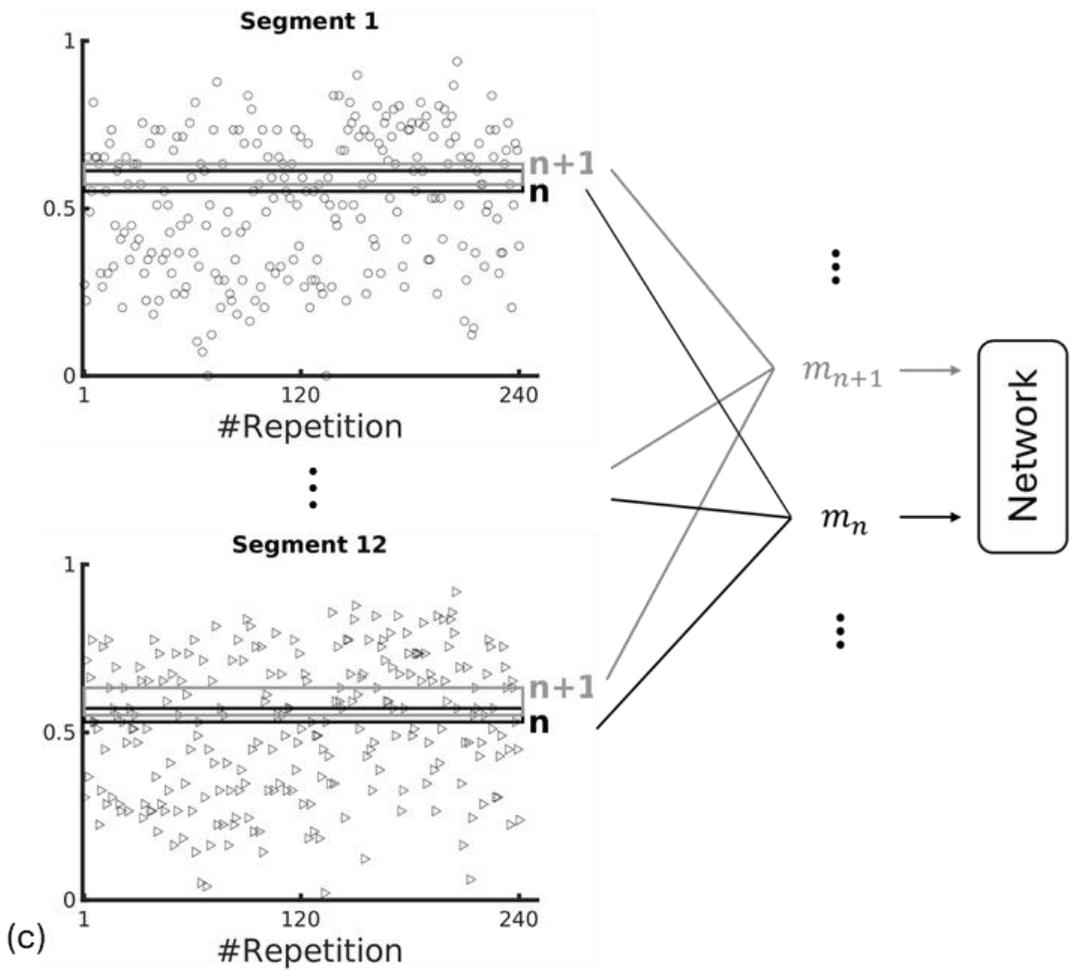

Figure 2: (a) Navigator signals from the selected main coil with normalized intensity. Local maxima (red motion curve) detected along the SI direction illustrates the respiratory motion of the liver. (b) The motion curve was normalized to 0 and 1, and the points in the motion curve were sorted into 12 groups based on the segment they belong to in their repetition. (c) A sliding window is applied to each segment group, and the motion signals in all groups bounded by the sliding windows with the same index are averaged. This average number is denoted as $m_n$ which will be used as the input of the 4D INR in the neural network as illustrated in figure 3. Sliding window n and n+1 have shared views.

## 2.3 | Image reconstruction

### 2.3.1| Optimization problem

Given the transient state image-series at reference-frame, the signal at echo i can be written as(23)

$$S_i(r) = U_r^H e^{-R_2^* t_i} e^{-i2\pi f t_i} (W' + F' e^{-i2\pi v_F t_i}) \quad (2)$$

for a single-peak fat model. In this equation, the terms W' and F' represent the singular images of water and fat, $v_F$ denotes the difference in precession frequency between water and fat. The variable $t_i$ represents the echo time $i$, and f stands for the total field. In our experiment, we specify the number of singular images used to be 4 as they together account for >99% of the energy of the simulated dictionary. Therefore, $W' = \begin{bmatrix} W_1' \\ W_2' \\ W_3' \\ W_4' \end{bmatrix}$ and $F' = \begin{bmatrix} F_1' \\ F_2' \\ F_3' \\ F_4' \end{bmatrix}$.

Recalling the strategy that we proposed to represent the motion states by a single number $m_n$ using sliding window-based method discussed in section 2.2 and shown in figure 2. The image series for motion state $m_n$ can be written as $S_{i,n} = U_r^H D_n e^{-R_2^* t_i} e^{-i2\pi f t_i}(W' + F' e^{-i2\pi v_F t_i})$, where $D_n$ is the motion deformation field for this state with respect to the reference-frame images. Then a NUFFT acting on $S_{i,n}$ should give the k-space data that matches the acquired k-space data associated with $m_n$. With this data consistency, reconstruction of the reference images and motion deformation field can be formulated as optimization problem:

$$(W', F', R_2^*, f, \{D_n\} \text{Error! Bookmark not defined. Error! Bookmark not defined.})$$
$$= argmin \sum_{n,i} \left| K_{n,i} - NUFFT\left(SU_r^H D_n e^{-R_2^* t_i} e^{-i2\pi f t_i}(W' + F' e^{-i2\pi v_F t_i})\right)\right|_1$$
$$+ \lambda ||\nabla D_n||_2^2 \quad (3)$$

where S is the coil sensitivity map and $K_{n,i}$ represents the k-space data of echo I associated with the $n_{th}$ sliding window. A total variation regularization is used on the motion deformation fields. By solving this optimization problem, singular images of water and fat for the reference-frame respiratory state, $\boldsymbol{W'}$ and $\boldsymbol{F'}$, motion deformation field associated with all respiratory states $\{D_n\}$, R2* map, and total field $f$ are given by the output of neural network. We propose a neural network using implicit neural representation (INR) based on the advanced multi-resolution hash encoding technique.(34)

### 2.3.2 | Network structure

Our network consists of two INRs, which were built using multilayer perceptrons (MLPs). Figure 3 outlines the workflow of the network. The INRs for the reference-frame image of MRI modalities and motion displacement field are depicted in the lower and upper branch of the network respectively. The INR for the MRI modalities ($f, R_2^*, \boldsymbol{W'}, \boldsymbol{F'}$) mapped 3D voxel coordinates to their corresponding image intensities. The input coordinates were pre-processed through a learning-based position encoding, $h_I$, before being fed into the MLPs to enhance the learning of high-frequency features. In this position encoding technique, feature vectors were associated with vertices of the grids of different resolutions and were trainable parameters.(34) They were encoded by a hash function and stored in a hash table. An input coordinate queries the feature of the image at a voxel by interpolating the feature vectors associated with adjacent grid vertices from multi-resolutions and combining these features vectors using concatenation, and thus the length of the queried feature depends on the number of resolution levels. In multi-parametric MRI, MRI modalities of the reference-frame images, $f, R_2^*, \boldsymbol{W'}, \boldsymbol{F'}$, share the same anatomy. Therefore, the learned multi-resolution feature vector by hash-encoding $h_I$ can be shared among the MLPs used to reconstruct these modalities:

$$f, R_2^*, \boldsymbol{W'}, \boldsymbol{F'} = MLP(h_I(\boldsymbol{x}, \boldsymbol{y}, \boldsymbol{z})) \quad (4)$$

The number of resolution levels we chose for the hash-table for $f, R_2^*, \boldsymbol{W'}, \boldsymbol{F'}$ reconstruction is 16, and the feature dimension of each level is 2, leading to a feature of size of 32 as the input of the MLPs.

As $\boldsymbol{W'}, \boldsymbol{F'}$ are complex images, there are two MLPs for the real and imaginary components for each of them. $R_2^*$ and $f$ are real images and thus are outputted by single MLPs. Each MLP consisted of an input layer, 4 hidden layers, and an output layer. The feature size for the input layer and hidden layers for all the above-mentioned MLPs are 32 and 64 respectively. As 4 singular images were used for the subspace recon, the feature size of the output layer of the MLPs for $\boldsymbol{W'}, \boldsymbol{F'}$ was 4, and the size of output layer of the MLPs for $R_2^*$ and $f$ was 1.

The upper branch of the network, for the estimation of the motion deformation field, takes the 3D voxel coordinate with a motion representation value, $m_n$ (equation 1), as input. The hash-encoding in this branch, denoted as $h_D$, outputs the feature of the motion deformation associated with $m_n$.

$$D_n = MLP(h_D(\boldsymbol{x}, \boldsymbol{y}, \boldsymbol{z}, m_n)) \quad (5)$$

We only use 5 levels of resolution for $h_D$, which helps the training convergence. The following MLP had an input layer, 3 hidden layers, and an output layer, and the sizes of these layers are 10, 32 and 3 respectively, as the final layer outputs the x,y,z components of the displacement field.

### 2.3.3 | Training details

For each iteration, a random sliding window index n was chosen by the data loader, which gives a groups of k-space readouts with motion state represented by $m_n$ bound by the sliding windows, calculated using equation (1). $m_n$ together with the 3D voxel-coordinate were the input of the network, and the above k-space data were used to calculate the data consistency loss (3) to update the network weights. One epoch traverses all the sliding window indices. With the help of the motion representation value $m_n$, the INR learns a smooth representation of the motion displacement field from end-of-expiration to end-of-inspiration.

Due to GPU memory limitation, 8-echo image volumes were not able to fit into the GPU all together. To resolve this issue, the training was divided into two phases: In phase one, only the first 4 echoes of the mGRE k-space are used for training. In phase two, echo 1 and 3 other echoes randomly selected from echo 2-8 were used to update the neural network. In this way, the network was trained for a total of 6 epochs; phase one contains the first two epochs, and phase two involves the rest 4 epochs. The network was implemented in pytorch using ADAM optimizer (initial learning rate being 0.01) on a NVIDIA A6000 GPU. TorchKbNUFFT, a pytorch based non-uniform library was used in this work.(36)The training time for one subject was approximately 3 hours.

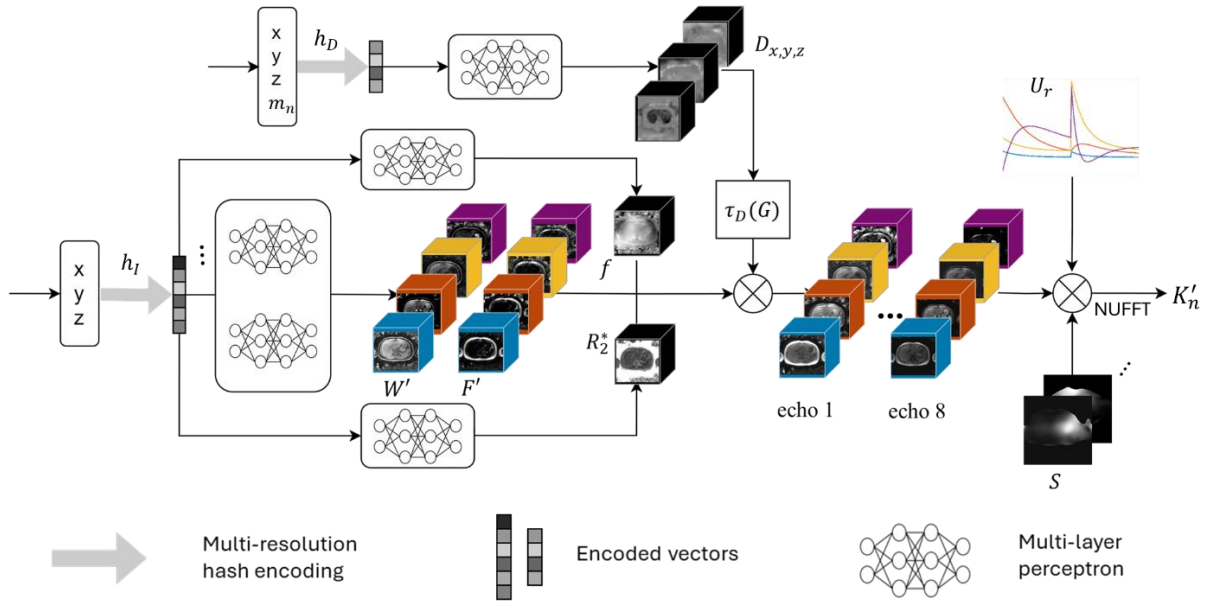

Figure 3: Network structure of the implicit neural representation-based MR fingerprinting (INR-MRF). The network simultaneously learns the reference frame images and the motion deformation field ($D_{x,y,z}$), which is enabled by a 3D INR and a 4D INR respectively. The reference frame images include water singular images ($W'$), fat singular images ($F'$), R2* map, and B0 field ($f$), which are the outputs of multiple parallel MLPs. The motion deformation field is used to register the reference frame images to the images with motion state represented by the input $m_n$. The deformed images are combined to generate the multi-echo complex images $D_n e^{-R_2^* t_i} e^{-i2\pi f t_i}(W' + F' e^{-i2\pi v_F t_i})$ (i =1,2,…,8). The training loss consists of a k-space data-consistency loss and a spatial total variation loss on the learned motion deformation fields. S: coil sensitivity map. $U_r$: principal components compressed from the dictionary. $\tau_D(G)$: deformed Cartesian grids.

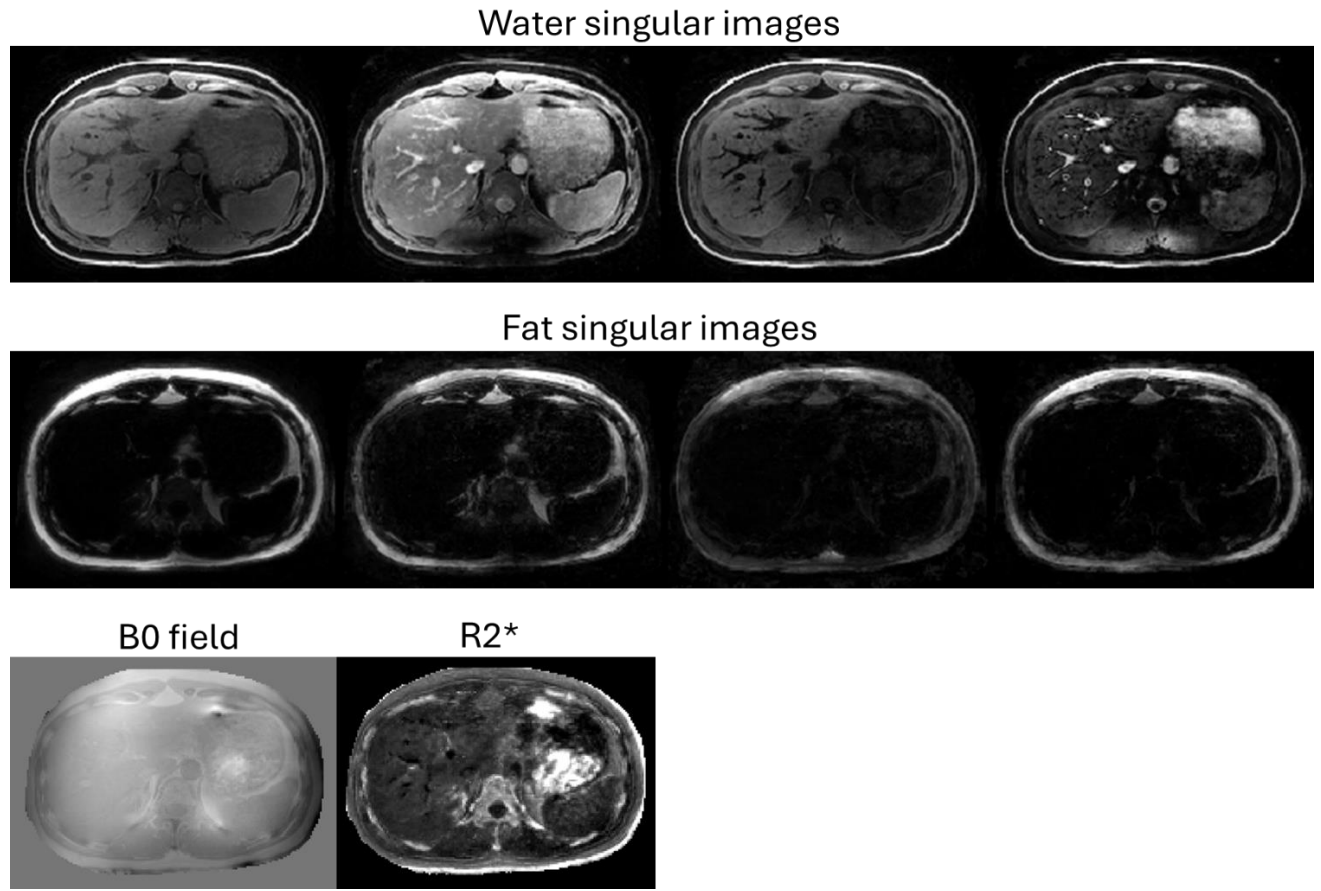

Figure 4: Water singular images, fat singular images, B0 field and R2* map of the healthy subject as output of the neural network.

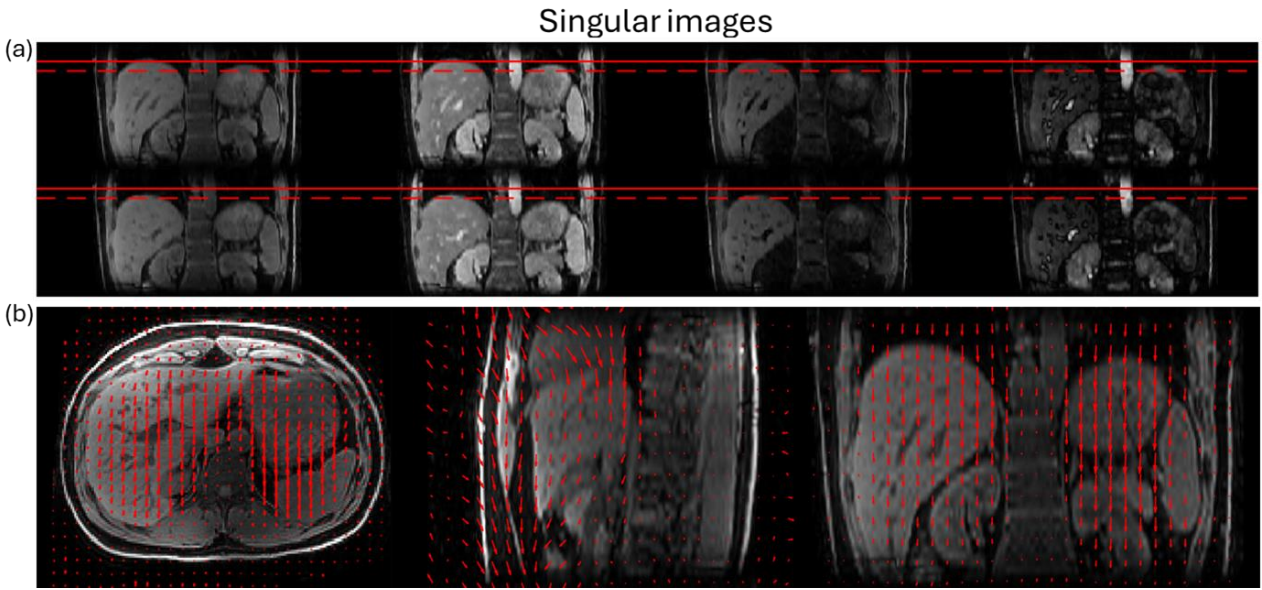

Figure 5: (a) Representative singular images from a healthy volunteer showing the liver at end-of-inspiration (top row) and end-of-expiration (bottom row). Solid red line indicates position of the boundary of the liver at end-of-inspiration, and the dashed red line indicates the position of end-of-expiration. (b) The motion deformation field learned by the neural network presented in vector form. The vectors in the axial view (left), sagittal view (middle) and coronal view (right) of the first singular image illustrate the direction of movement from end-of-inspiration to end-of-expiration.

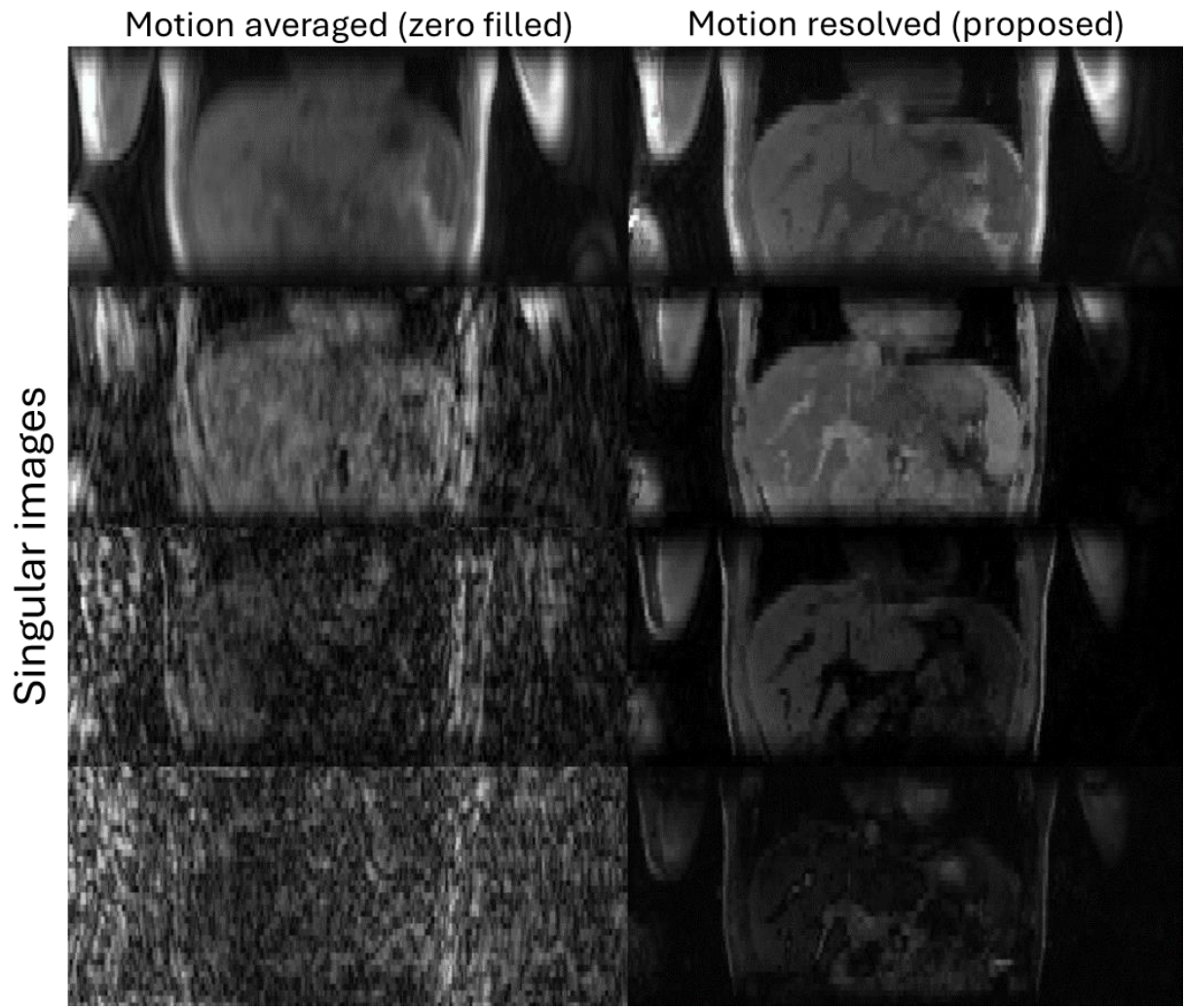

Figure 6: Singular images reconstructed by zero-filling the k-space versus by INR-MRF. Substantial motion artifacts and aliasing artifacts were suppressed.

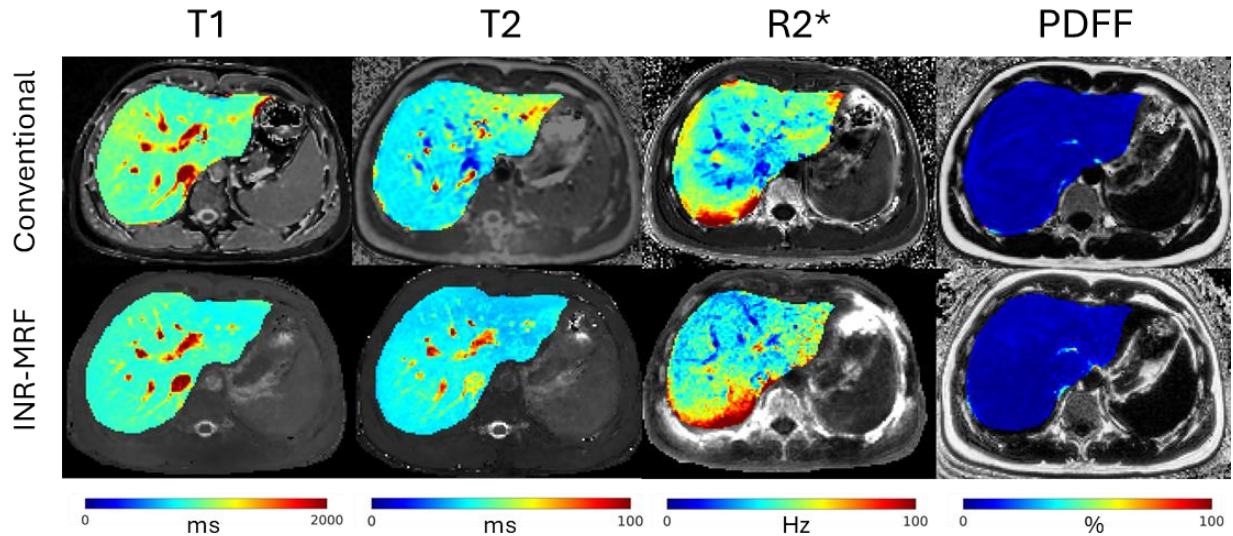

Figure 7: Representative T1, T2, R2* and PDFF maps from a healthy volunteer produced by conventional methods (first row) and INR-MRF (second row). Note that T1 and T2 maps generated by INR-MRF are water-specific.

## 2.4 | Data acquisition and post processing

The experiments were conducted on 10 clinically healthy subjects (5 females; age 28 ± 3.5 years) with no history of liver disease. This experiment received approval from the local institutional review board, and all subjects gave written informed consent prior to scanning. The experiments were performed using a 3T scanner (GE Healthcare, Waukesha, WI, USA), employing a 16-channel array coil. For reference, a 4-echo Cartesian multi-echo gradient-echo sequence was used to produce the water, fat and R2*images in a breath-hold at end-expiration with the following parameters: FOV = 480 × 480 mm², reconstruction matrix size = 256 × 256 x 20, TE = 2.8/6.1/9.4/12.7 ms, TR = 15.82 ms, and slice thickness = 5 mm. 2D MOLLI and 2D multi-echo spin echo (MESE) were acquired during breath-hold at end-expiration for 1 slice, taken from the middle of the liver. The parameters for MOLLI were: FOV = 380 mm², matrix = 256x256, and slice thickness = 5 mm, TE = 1.29, TR = 2.97, and the parameters for MESE were: FOV = 380 mm², matrix = 256x256, and slice thickness = 5 mm, TR = 350 ms, TE = 6.25/12.50/18.75/25/31.25/37.50/43.75/50 ms.  The proposed INR-MRF sequence was executed, covering the entire liver with parameters: FOV = 480 × 480 mm², matrix = 256 × 256 mm², TE = 0.54 /1.70/2.87/4.03/5.19/6.36/7.52/8.68ms, TR = 14.2 ms, slice thickness = 5 mm, and number of slices = 40. For more information about the pulse sequences, please refer to Table S1.

T1 and T2 maps for INR-MRF were determined using a dictionary matching approach. A dictionary of transverse magnetization for the INR-MRF sequence was created via a numerical

Bloch simulation, using the sequence parameters outlined in Table S1 and assuming an on-resonance condition. The T1 values in [100:2:1000, 1000:10:2000], and T2 values in [10:1:100, 100:5:1000] were used to generate the dictionary. The Proton Density Fat Fraction (PDFF) for INR-MRF is estimated using the first water and fat singular images ($W_1'$ and $F_1'$), calculated as $F_1'/(W_1' + F_1')$.

For the reference Cartesian mGRE, an initial estimation of the field f was produced from the complex mGRE signals, achieved through concurrent phase unwrapping and chemical shift elimination (SPURS)(37). The IDEAL algorithm (38) was then applied to create maps of water ($W$), fat ($F$), and field ($f$) mGRE data PDFF were generated from the computed fat and water images as $F/(W + F)$. The reference R2* and T2 map was obtained using the ARLO algorithm(39) with mGRE and mSE data respectively. Reference T1 map was generated by fitting the MOLLI data to the exponential model described in (19) using *fitnlm* function in MATLAB (2023b).

## 2.5 | Analysis

ROIs for T1 and T2 map generated by our method were manually drawn on the slice corresponding to the slice of MOLLI image and MESE image, and ROIs for R2* and PDFF were drawn on the 3D volume corresponding to the 3D Cartesian mGRE images, including as much liver tissue as possible while avoiding blood vessels and edges. The mean values of T1, T2, R2* and PDFF in the liver ROIs for INR-MRF and conventional scans were compared. The repeatability of the measurements was assessed using Bland-Altman plots.

## 3 | Results

Reference frame water and fat singular images, R2* and B0 field, as outputs of the neural network, were shown in figure 4, revealing sharp image quality with abundant details for liver tissues and veins. The reference frame images registered to the end-of-inspiration and end-of-expiration state of the liver (figure 5.a) using deformation fields learned by the network were illustrated, demonstrating the ability of INR-MRF to resolve respiratory motions. Singular images generated by INR-MRF were compared with zero-filled reconstruction in figure 6 where substantial blurring, aliasing and motion artifacts were reduced. T1, T2, R2* and PDFF maps generated using INR-MRF for a representative subject were compared with the conventional methods in figure 7, showing good consistency across the liver.

Figure 8 shows Bland–Altman plots of average liver T1, T2, R2* and PDFF values obtained from the 10 subjects, demonstrating small bias and narrow 95% limits of agreement. The T1 value of the liver measured using MOLLI was $921.7 \pm 28.8$ ms, and that measured by INR-MRF was $884.6 \pm 41.9$ ms. Liver T2 values measured by MESE and INR-MRF were $37.9 \pm 2.8$ ms and $35.8 \pm 4.5$ ms respectively. R2* and PDFF evaluated using Cartesian mGRE sequence were $47.9 \pm 7.9$ Hz and $4.5 \pm 1.2$ % respectively, and R2* and PDFF evaluated using INR-MRF were $49.2 \pm 9.8$ Hz and $4.9 \pm 1.8$ % respectively.

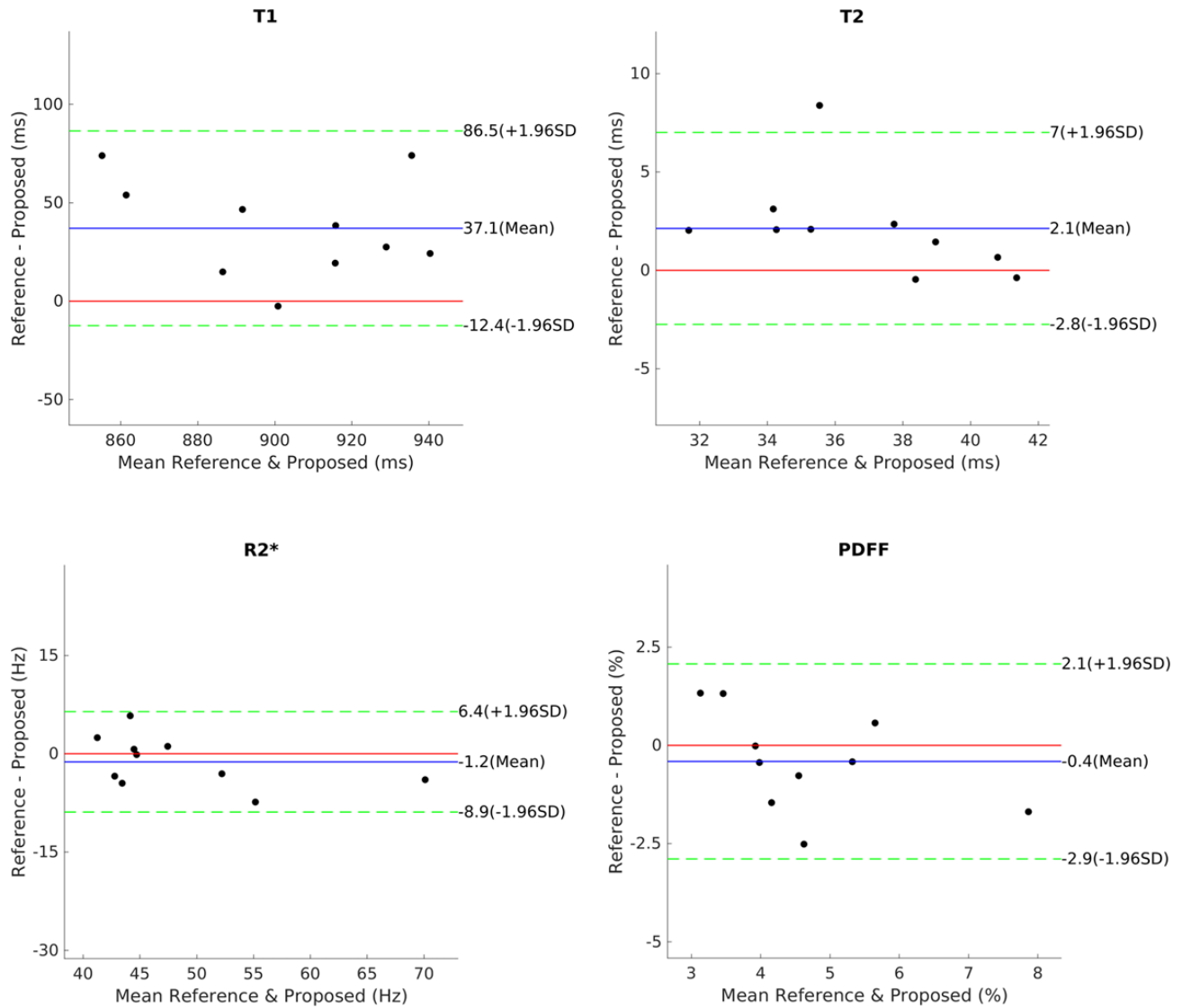

Figure 8: Bland–Altman plots of average T1, T2, R2* and PDFF values for the liver obtained with the proposed INR-MRF and reference methods from 10 healthy subjects. Small bias and narrow 95% limits of agreement were achieved by INR-MRF.

## 4. Discussion

In this study, we introduced INR-MRF, an unsupervised learning framework that jointly learns the reference frame MRI images and motion deformation fields in zero-shot. INR-MRF utilizes a 3D and a 4D INR, along with a registration module to produce dynamic MR images with enhanced image quality. In addition, the network simultaneously separates water and fat images in the training process, eliminating the need of performing any retrospective water-fat separation.

We proposed a contrast-selective sliding window-based motion binning pipeline. Using the motion representation value extracted from the binned navigator signals and 3D Cartesian coordinate as input, the network learns the water and fat singular images, R2* map, B0 field, and motion deformation fields throughout the respiratory cycle. T1 and T2 maps can be generated using dictionary matching. Our in-vivo experiment showed comparable quantitative values of INR-MRF in liver with respect to the reference scans.

Our method poses several limitations. First, in our model, the B0 field being subject to the same motion deformation fields as the magnitude images is an approximation. B0 field consists of tissue local field and background field, which are convolution of a magnetic dipole kernel with tissue local susceptibility sources and susceptibility sources from the background (e.g. air) respectively. The component that deforms in the way described by the learned motion deformation fields was the local field, or more precisely, the local susceptibility sources. As tissue susceptibility contains rich information about the pathological conditions of the livers, e.g. quantitative susceptibility mapping (QSM)(40, 41) has been used to diagnose liver iron deposition and evaluate fibrosis,(42-44) the B0 field obtained in our method might not be very useful to reconstruct an accurate susceptibility map. To get more comprehensive information of the liver including the susceptibility map, a network that better models the physical process needs to be established.

The pulse sequence design was also preliminary, as there was only one inversion recovery pulse and one T2 preparation pulse included. The T2 prep echo time, 37.5 ms, used in our pulse sequence was designed for the liver, which might be too short for other abdominal organs with longer T2. Luckily, it was found that a T2 prep echo time ~85ms was enough to capture a large range of T2 from 50ms to 150ms with negligible errors compared with the ground truth.(26) Our sequence can be easily extended to measure the whole abdomen quantitative maps with a simply increasing the T2 prep echo time. On the other hand, more IR and T2 prepared acquisition blocks can be added to the pulse sequence as proposed by (22, 23). As long as the principal components compressed from the corresponding dictionary are calculated, our deep learning reconstruction framework can be used on any pulse sequence designs.

## 5. Conclusion

In this work, we demonstrated the feasibility of acquiring 3D co-registered whole liver T1, T2, R2* and PDFF maps in vivo in a free-running scan. Our proposed deep learning frame utilizes implicit neural representation to jointly learns the reference frame MRI images and motion deformation fields in zero-shot.

## Supporting information

Table S1. Pulse sequence parameters of the proposed method (INR-MRF) and reference scans (Cartesian mGRE, MESE, MOLLI)

|  | INR-MRF (proposed) | mGRE | MESE | MOLLI |
|---|---|---|---|---|
| FOV(mm$^2$) | 480 x 480 | 480 x 480 | 380 x 380 | 380 x 380 |
| Acquisition dimension | 3D | 3D | 2D | 2D |
| Trajectory | Spiral | Cartesian | Cartesian | Cartesian |
| Matrix size | 256 x 256 | 256 x 256 | 256 x 256 | 256 x 256 |
| In-plane resolution (mm$^2$) | 1.9 x 1.9 | 1.9 x 1.9 | 1.5 x 1.5 | 1.5 x 1.5 |
| Slice thickness (mm) | 5 | 5 | 5 | 5 |
| Number of slices | 40 | 20 | 1 | 1 |
| TI (ms) | 300 | / | / | 227 |
| TR (ms) | 12.7 | 15.9 | 350 | 2.97 |
| TE (ms) | 0.5 | 9.4 | 12.5 | 1.29 |
| T2prep echo time (ms) | 37.5 | / | / | / |
| ΔTE | 1.2 | 3.3 | 6.3 | / |
| Flip angle (°) | 10 | 15 | 90 | 35 |

| Scan duration | 10-min free-breathing | 27-sec breath hold | 22-sec breath hold | 12-sec breath-hold |
|---|---|---|---|---|

1.	Moon AM, Singal AG, Tapper EB. Contemporary Epidemiology of Chronic Liver Disease and Cirrhosis. Clin Gastroenterol Hepatol. 2020;18(12):2650-66. doi: 10.1016/j.cgh.2019.07.060. PubMed PMID: 31401364; PMCID: PMC7007353.
2.	Younossi ZM, Wong G, Anstee QM, Henry L. The Global Burden of Liver Disease. Clin Gastroenterol Hepatol. 2023;21(8):1978-91. doi: 10.1016/j.cgh.2023.04.015. PubMed PMID: 37121527.
3.	Younossi Z, Tacke F, Arrese M, Chander Sharma B, Mostafa I, Bugianesi E, Wai-Sun Wong V, Yilmaz Y, George J, Fan J, Vos MB. Global Perspectives on Nonalcoholic Fatty Liver Disease and Nonalcoholic Steatohepatitis. Hepatology. 2019;69(6):2672-82. doi: 10.1002/hep.30251. PubMed PMID: 30179269.
4.	Schuppan D, Afdhal NH. Liver cirrhosis. Lancet. 2008;371(9615):838-51. doi: 10.1016/S0140-6736(08)60383-9. PubMed PMID: 18328931; PMCID: PMC2271178.
5.	Tsochatzis EA, Bosch J, Burroughs AK. Liver cirrhosis. Lancet. 2014;383(9930):1749-61. doi: 10.1016/S0140-6736(14)60121-5. PubMed PMID: 24480518.
6.	Adams PC, Jeffrey G, Ryan J. Haemochromatosis. Lancet. 2023;401(10390):1811-21. Epub 20230427. doi: 10.1016/s0140-6736(23)00287-8. PubMed PMID: 37121243.
7.	Corradini E, Pietrangelo A. Iron and steatohepatitis. J Gastroenterol Hepatol. 2012;27 Suppl 2:42-6. doi: 10.1111/j.1440-1746.2011.07014.x. PubMed PMID: 22320915.
8.	Hilton C, Sabaratnam R, Drakesmith H, Karpe F. Iron, glucose and fat metabolism and obesity: an intertwined relationship. Int J Obes (Lond). 2023;47(7):554-63. Epub 20230407. doi: 10.1038/s41366-023-01299-0. PubMed PMID: 37029208; PMCID: PMC10299911.
9.	Kouroumalis E, Tsomidis I, Voumvouraki A. Iron as a therapeutic target in chronic liver disease. World J Gastroenterol. 2023;29(4):616-55. doi: 10.3748/wjg.v29.i4.616. PubMed PMID: 36742167; PMCID: PMC9896614.
10.	Reeder SB, Starekova J. MRI Proton Density Fat Fraction for Liver Disease Risk Assessment: A Call for Clinical Implementation. Radiology. 2023;309(1):e232552. doi: 10.1148/radiol.232552. PubMed PMID: 37874237.
11.	Yokoo T, Bydder M, Hamilton G, Middleton MS, Gamst AC, Wolfson T, Hassanein T, Patton HM, Lavine JE, Schwimmer JB, Sirlin CB. Nonalcoholic fatty liver disease: diagnostic and fat-grading accuracy of low-flip-angle multiecho gradient-recalled-echo MR imaging at 1.5 T. Radiology. 2009;251(1):67-76. doi: 10.1148/radiol.2511080666. PubMed PMID: 19221054; PMCID: PMC2663579.
12.	Zhong X, Nickel MD, Kannengiesser SA, Dale BM, Kiefer B, Bashir MR. Liver fat quantification using a multi-step adaptive fitting approach with multi-echo GRE imaging. Magn Reson Med. 2014;72(5):1353-65. doi: 10.1002/mrm.25054. PubMed PMID: 24323332.
13.	Hernando D, Levin YS, Sirlin CB, Reeder SB. Quantification of liver iron with MRI: state of the art and remaining challenges. J Magn Reson Imaging. 2014;40(5):1003-21. doi: 10.1002/jmri.24584. PubMed PMID: 24585403; PMCID: PMC4308740.
14.	Wood JC, Enriquez C, Ghugre N, Tyzka JM, Carson S, Nelson MD, Coates TD. MRI R2 and R2* mapping accurately estimates hepatic iron concentration in transfusion-dependent thalassemia and sickle cell disease patients. Blood. 2005;106(4):1460-5. doi: 10.1182/blood-2004-10-3982. PubMed PMID: 15860670; PMCID: PMC1895207.

15.	Hoffman DH, Ayoola A, Nickel D, Han F, Chandarana H, Shanbhogue KP. T1 mapping, T2 mapping and MR elastography of the liver for detection and staging of liver fibrosis. Abdom Radiol (NY). 2020;45(3):692-700. doi: 10.1007/s00261-019-02382-9. PubMed PMID: 31875241.
16.	Guimaraes AR, Siqueira L, Uppal R, Alford J, Fuchs BC, Yamada S, Tanabe K, Chung RT, Lauwers G, Chew ML, Boland GW, Sahani DV, Vangel M, Hahn PF, Caravan P. T2 relaxation time is related to liver fibrosis severity. Quant Imaging Med Surg. 2016;6(2):103-14. doi: 10.21037/qims.2016.03.02. PubMed PMID: 27190762; PMCID: PMC4858462.
17.	Lee MJ, Kim MJ, Yoon CS, Han SJ, Park YN. Evaluation of liver fibrosis with T2 relaxation time in infants with cholestasis: comparison with normal controls. Pediatr Radiol. 2011;41(3):350-4. doi: 10.1007/s00247-010-1874-5. PubMed PMID: 20959973.
18.	Hoad CL, Palaniyappan N, Kaye P, Chernova Y, James MW, Costigan C, Austin A, Marciani L, Gowland PA, Guha IN, Francis ST, Aithal GP. A study of T(1) relaxation time as a measure of liver fibrosis and the influence of confounding histological factors. Nmr Biomed. 2015;28(6):706-14. doi: 10.1002/nbm.3299. PubMed PMID: 25908098.
19.	Messroghli DR, Radjenovic A, Kozerke S, Higgins DM, Sivananthan MU, Ridgway JP. Modified Look-Locker inversion recovery (MOLLI) for high-resolution T1 mapping of the heart. Magn Reson Med. 2004;52(1):141-6. doi: 10.1002/mrm.20110. PubMed PMID: 15236377.
20.	Fatemi Y, Danyali H, Helfroush MS, Amiri H. Fast T(2) mapping using multi-echo spin-echo MRI: A linear order approach. Magn Reson Med. 2020;84(5):2815-30. doi: 10.1002/mrm.28309. PubMed PMID: 32430979; PMCID: PMC7402028.
21.	Ma D, Gulani V, Seiberlich N, Liu K, Sunshine JL, Duerk JL, Griswold MA. Magnetic resonance fingerprinting. Nature. 2013;495(7440):187-92. doi: 10.1038/nature11971. PubMed PMID: 23486058; PMCID: PMC3602925.
22.	Cruz G, Qi H, Jaubert O, Kuestner T, Schneider T, Botnar RM, Prieto C. Generalized low-rank nonrigid motion-corrected reconstruction for MR fingerprinting. Magn Reson Med. 2022;87(2):746-63. doi: 10.1002/mrm.29027. PubMed PMID: 34601737.
23.	Jaubert O, Arrieta C, Cruz G, Bustin A, Schneider T, Georgiopoulos G, Masci PG, Sing-Long C, Botnar RM, Prieto C. Multi-parametric liver tissue characterization using MR fingerprinting: Simultaneous T(1) , T(2) , T(2) *, and fat fraction mapping. Magn Reson Med. 2020;84(5):2625-35. doi: 10.1002/mrm.28311. PubMed PMID: 32406125.
24.	Heesterbeek DGJ, Koolstra K, van Osch MJP, van Gijzen MB, Vos FM, Nagtegaal MA. Mitigating undersampling errors in MR fingerprinting by sequence optimization. Magn Reson Med. 2023;89(5):2076-87. doi: 10.1002/mrm.29554. PubMed PMID: 36458688.
25.	Bustin A, Lima da Cruz G, Jaubert O, Lopez K, Botnar RM, Prieto C. High-dimensionality undersampled patch-based reconstruction (HD-PROST) for accelerated multi-contrast MRI. Magn Reson Med. 2019;81(6):3705-19. doi: 10.1002/mrm.27694. PubMed PMID: 30834594; PMCID: PMC6646908.
26.	Zhang J, Nguyen TD, Solomon E, Li C, Zhang Q, Li J, Zhang H, Spincemaille P, Wang Y. mcLARO: Multi-contrast learned acquisition and reconstruction optimization for simultaneous quantitative multi-parametric mapping. Magn Reson Med. 2024;91(1):344-56. doi: 10.1002/mrm.29854. PubMed PMID: 37655444.
27.	Wang N, Cao T, Han F, Xie Y, Zhong X, Ma S, Kwan A, Fan Z, Han H, Bi X, Noureddin M, Deshpande V, Christodoulou AG, Li D. Free-breathing multitasking multi-echo MRI for whole-liver water-specific T(1) , proton density fat fraction, and R2 * quantification. Magn Reson Med. 2022;87(1):120-37. doi: 10.1002/mrm.28970. PubMed PMID: 34418152; PMCID: PMC8616772.
28.	Byra M, Poon C, Rachmadi MF, Schlachter M, Skibbe H. Exploring the performance of implicit neural representations for brain image registration. Sci Rep. 2023;13(1):17334. doi: 10.1038/s41598-023-44517-5. PubMed PMID: 37833464; PMCID: PMC10575995.

29. Han X, Chen Z, Lin G, Lv W, Zheng C, Lu W, Sun Y, Lu L. Semi-supervised model based on implicit neural representation and mutual learning (SIMN) for multi-center nasopharyngeal carcinoma segmentation on MRI. Comput Biol Med. 2024;175:108368. doi: 10.1016/j.compbiomed.2024.108368. PubMed PMID: 38663351.
30. Liu L, Shen L, Johansson A, Balter JM, Cao Y, Vitzthum L, Xing L. Volumetric MRI with sparse sampling for MR-guided 3D motion tracking via sparse prior-augmented implicit neural representation learning. Med Phys. 2024;51(4):2526-37. doi: 10.1002/mp.16845. PubMed PMID: 38014764; PMCID: PMC10994763.
31. Shao HC, Mengke T, Deng J, Zhang Y. 3D cine-magnetic resonance imaging using spatial and temporal implicit neural representation learning (STINR-MR). Phys Med Biol. 2024;69(9). doi: 10.1088/1361-6560/ad33b7. PubMed PMID: 38479004; PMCID: PMC11017162.
32. Wu Q, Li Y, Sun Y, Zhou Y, Wei H, Yu J, Zhang Y. An Arbitrary Scale Super-Resolution Approach for 3D MR Images via Implicit Neural Representation. IEEE J Biomed Health Inform. 2023;27(2):1004-15. doi: 10.1109/JBHI.2022.3223106. PubMed PMID: 37022393.
33. Sitzmann V, Martel JNP, Bergman AW, Lindell DB, Wetzstein G. Implicit Neural Representations with Periodic Activation Functions. Adv Neur In. 2020;33. PubMed PMID: WOS:001207690600066.
34. Müller T, Evans A, Schied C, Keller A. Instant Neural Graphics Primitives with a Multiresolution Hash Encoding. Acm T Graphic. 2022;41(4). doi: Artn 102

10.1145/3528223.3530127. PubMed PMID: WOS:000830989200074.
35. Feng L, Axel L, Chandarana H, Block KT, Sodickson DK, Otazo R. XD-GRASP: Golden-angle radial MRI with reconstruction of extra motion-state dimensions using compressed sensing. Magn Reson Med. 2016;75(2):775-88. doi: 10.1002/mrm.25665. PubMed PMID: 25809847; PMCID: PMC4583338.
36. M. J. Muckley RS, T. Murrell and F. Knoll, editor. TorchKbNufft: A High-Level, Hardware-Agnostic Non-Uniform Fast Fourier Transform. ISMRM Workshop on Data Sampling & Image Reconstruction; 2020.
37. Dong J, Liu T, Chen F, Zhou D, Dimov A, Raj A, Cheng Q, Spincemaille P, Wang Y. Simultaneous phase unwrapping and removal of chemical shift (SPURS) using graph cuts: application in quantitative susceptibility mapping. IEEE Trans Med Imaging. 2015;34(2):531-40. doi: 10.1109/TMI.2014.2361764. PubMed PMID: 25312917.
38. Reeder SB, McKenzie CA, Pineda AR, Yu H, Shimakawa A, Brau AC, Hargreaves BA, Gold GE, Brittain JH. Water-fat separation with IDEAL gradient-echo imaging. J Magn Reson Imaging. 2007;25(3):644-52. doi: 10.1002/jmri.20831. PubMed PMID: 17326087.
39. Pei M, Nguyen TD, Thimmappa ND, Salustri C, Dong F, Cooper MA, Li J, Prince MR, Wang Y. Algorithm for fast monoexponential fitting based on Auto-Regression on Linear Operations (ARLO) of data. Magn Reson Med. 2015;73(2):843-50. doi: 10.1002/mrm.25137. PubMed PMID: 24664497; PMCID: PMC4175304.
40. Wang Y, Liu T. Quantitative susceptibility mapping (QSM): Decoding MRI data for a tissue magnetic biomarker. Magn Reson Med. 2015;73(1):82-101. doi: 10.1002/mrm.25358. PubMed PMID: 25044035; PMCID: PMC4297605.
41. Liu J, Liu T, de Rochefort L, Ledoux J, Khalidov I, Chen WW, Tsiouris AJ, Wisnieff C, Spincemaille P, Prince MR, Wang Y. Morphology enabled dipole inversion for quantitative susceptibility mapping using structural consistency between the magnitude image and the susceptibility map. Neuroimage. 2012;59(3):2560-8. doi: 10.1016/j.neuroimage.2011.08.082. PubMed PMID: WOS:000299494000057.
42. Li C, Zhang, J., Dimov, A.V., de González, A.K.K., Prince, M.R., Li, J., Romano, D., Spincemaille, P., Nguyen, T.D., Brittenham, G.M. and Wang, Y. MRI quantification of liver fibrosis using diamagnetic susceptibility: An ex-vivo feasibility study. arXiv preprint arXiv:2410.031272024.

43.	Yoshikawa M, Kudo K, Harada T, Harashima K, Suzuki J, Ogawa K, Fujiwara T, Nishida M, Sato R, Shirai T, Bito Y. Quantitative Susceptibility Mapping versus R2*-based Histogram Analysis for Evaluating Liver Fibrosis: Preliminary Results. Magn Reson Med Sci. 2022;21(4):609-22. doi: 10.2463/mrms.mp.2020-0175. PubMed PMID: WOS:000882515800001.
44.	Jafari R, Hectors SJ, Koehne de Gonzalez AK, Spincemaille P, Prince MR, Brittenham GM, Wang Y. Integrated quantitative susceptibility and R(2) * mapping for evaluation of liver fibrosis: An ex vivo feasibility study. Nmr Biomed. 2021;34(1):e4412. doi: 10.1002/nbm.4412. PubMed PMID: 32959425; PMCID: PMC7768551.